\documentclass[aip,apl,reprint,amsmath,amssymb]{revtex4-1}
\usepackage[utf8]{inputenc}
\usepackage[T1]{fontenc}
\usepackage[english]{babel}

\usepackage{dcolumn}
\usepackage{multirow}
\usepackage{graphicx}
\graphicspath{{pic/}}

\usepackage{hyperref}
\hypersetup{colorlinks=true,linkcolor=blue,citecolor=magenta}

\begin{document}
%\preprint{Article about particle lofting. Draft. \today}

\title{Particle lofting from substrate exposed to plasma and electron beam} %Title of paper

\author{P.~V.~Krainov}
\email{pavel.krainov@phystech.edu}
\affiliation{Institute for Spectroscopy of the Russian Academy of Sciences, 
Fizicheskaya str. 5, Troitsk, Moscow 108840, Russia}
\affiliation{Moscow Institute of Physics and Technology, 
Institutskiy pereulok str. 9, Dolgoprudny, Moscow region 141701, Russia}

\author{V.~V.~Ivanov}
\affiliation{Institute for Spectroscopy of the Russian Academy of Sciences, 
Fizicheskaya str. 5, Troitsk, Moscow 108840, Russia}

\author{D.~I.~Astakhov}
\affiliation{Institute for Spectroscopy of the Russian Academy of Sciences, 
Fizicheskaya str. 5, Troitsk, Moscow 108840, Russia}
\affiliation{ISTEQ B.V., High Tech Campus 9, 5656 AE Eindhoven, The Netherlands}

\author{V.~V.~Medvedev}
\affiliation{Institute for Spectroscopy of the Russian Academy of Sciences, 
Fizicheskaya str. 5, Troitsk, Moscow 108840, Russia}
\affiliation{Moscow Institute of Physics and Technology, 
Institutskiy pereulok str. 9, Dolgoprudny, Moscow region 141701, Russia}

\author{V.~V.~Kvon}
\affiliation{ASML Netherlands B.V., De Run 6501, 5504DR Veldhoven, The Netherlands}

\author{A.~M.~Yakunin}
\affiliation{ASML Netherlands B.V., De Run 6501, 5504DR Veldhoven, The Netherlands}

\author{M.~A.~van~de~Kerkhof}
\affiliation{ASML Netherlands B.V., De Run 6501, 5504DR Veldhoven, The Netherlands}

\date{\today}

\begin{abstract}
A nanometer-sized dielectric particle lying on a dielectric substrate 
is exposed to the flux of low-energy electrons,
ion and electron fluxes from a cold plasma 
and the fluxes from the combination of these two sources with the help of particle-in-cell simulation
to investigate the particle lofting phenomenon.
The results are of interest for dust mitigation in the semiconductor industry,
the lunar exploration, and the explanation of the dust levitation.
\end{abstract}


\maketitle

\section{Introduction\label{sec:intro}}
EUVL scanners \cite{vandekerkhof_2019} operate in a low-pressure hydrogen atmosphere \cite{beckers_2019}. 
The propagation of ionizing EUV radiation in this atmosphere inevitably leads to 
the formation of hydrogen plasma. 
In turn, this plasma interacts with the walls of the scanner chamber and with various functional surfaces. 
In this case, exposure to plasma can lead to the release of pollution particles from such surfaces and 
their further transport to the sensitive surfaces of the mask and wafer \cite{vandekerkhof_2019a,scaccabarozzi_2009}. 
Controlling and minimizing particle contamination requires detailed investigation of 
the particle release mechanisms.
 
Experimental and theoretical studies of the release of particles from surfaces under the influence of plasma 
have been carried out since the early 1990s. 
These studies were aimed at interpreting the observation of levitation of dust particles 
over the surface of airless space bodies such as the Moon or asteroids. 
Experimental studies were carried out mainly using micron-sized regolith particles. 
Nonetheless, these published results allow the theoretical particle release models to be tested and validated.

Several experimental groups detected the most intense particle release by simultaneous exposure to 
low-temperature plasma and low-energy electron beam \cite{sheridan_1992,flanagan_2006,wang_2016}.
Besides, some of them detected release of dust particles exposed to only electron beam \cite{wang_2016}.
In the recent work \cite{krainov_2020a} we studied lofting by electron beam with the help of 
particle-in-cell numerical simulation.
We found that interaction between charges accumulated on surfaces of particle and substrate 
can be the reason of lofting. 
This finding is in agreement with the recently developed "patched charge model" \cite{wang_2016,schwan_2017a}
but in contrast to the widespread hypothesis \cite{flanagan_2006,sheridan_2011} 
that explains particles transport by interaction with the electric field of plasma sheath.

In this work we apply our particle-in-cell model \cite{astakhov_2016} for further investigation of 
the lofting phenomenon.
We carried out a simulation of nanometer-sized dielectric particle lofting resting on 
a dielectric substrate in a chamber filled by hydrogen and exposed to low-energy electron beam.
By varying hydrogen pressure, we investigated a range of conditions from a pure electron beam 
to a low-temperature plasma including their combination. 
We found that a plasma addition to an electron beam increases repulsive force 
and can make particle lofting possible from a dielectric substrate as thin as a naturally formed oxide layer.

The simulation consists of two stages.
The first stage described in sec.~\ref{sec:plasma_formation} 
is the simulation of a plasma formation in a chamber filled by hydrogen 
and induced by an electron beam.
The second stage described in sec.~\ref{sec:buildup} is the simulation of a particle charge buildup under 
the exposure of ions and electrons from plasma.

\section{Forces\label{sec:forces}}
The object of our study is a dielectric particle lying on a dielectric surface.
The particle experiences the action of five forces.
Two repulsive electrostatic forces: 
the interaction with an external electric field (e.g., field of plasma sheath) 
and the interaction between charges on the particle surface and charges on the substrate surface,
two attractive forces: a surface adhesive force 
and the interaction of the particle charge with its electric image in the dielectric substrate (mirror force),
and gravitational force.

It was concluded previously in several papers\cite{sheridan_1992,flanagan_2006,krainov_2020a}
that gravitational force is several orders lower than adhesive van der Waals force.
Therefore, there are only four forces to be considered.

As a major surface adhesive force the van der Waals force is considered. 
In spite of the availability of the Hamaker theory and several more comprehensive theories\cite{lamarche_2017}
for estimation of van der Waals force, its value is not a well defined.
A literature review gives a range from $10^{-12}$ to $10^{-8}$~N for 100~nm particle.

The lower limit of this range requires $n\cdot E \geq 10^{7}$~$Vm^{-1}$,
where $n = |Q/e|$ is the number of charges collected by the particle and $E$ is the electric field.
This requires a supercharge of particle or extremely large value of electric field,
that is not reached in a plasma used in any conducted experiment about particle lofting.

Recently we demonstrated that other two forces is of great importance for particle lofting 
in case of exposure to low-energy electric beam.
The calculation of other two forces in case of exposure to plasma and electron beam 
is the issue discussed in this article.

\section{Simulation of plasma formation\label{sec:plasma_formation}}
\begin{figure}[ht]
  \includegraphics[width=\linewidth]{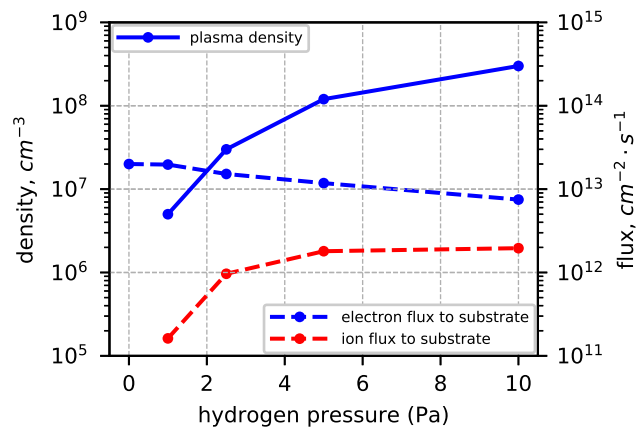}
  \caption{The dependencies of plasma density and fluxes to the substrate on hydrogen pressure.  
  \label{fig:plasma_density}}
\end{figure}

The simulation of plasma formation was conducted with the help of
a two-dimensional cylindrically symmetrical particle-in-cell (PIC) model \cite{astakhov_2016},
that follows the general PIC scheme \cite{charleskbirdsall_1991}.
At every time step ions and electrons represented by super particles update their positions and velocities 
in accordance with electric field distribution.
Then, a Poisson equation solver is used to obtain a new electric field distribution 
corresponding to the distribution of ions and electrons. 
A Monte-Carlo (MC) scheme was used to introduce collisions of electrons, ions and fast neutrals with molecules 
from the background gas \cite{astakhov_2016}.
Electron-electron, electron-ion, ion-ion Coulomb collisions as well as three body collision were neglected,
because of low probability due to low plasma density (fig.~\ref{fig:plasma_density}).
The model was previously validated in experiments 
with extreme ultraviolet induced hydrogen plasma \cite{astakhov_2015,astakhov_2016a,abrikosov_2017}.

\begin{figure}[ht]
  \includegraphics[width=\linewidth]{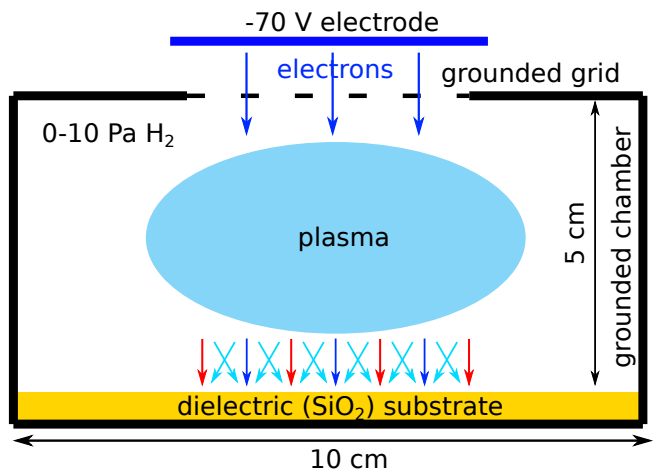}
  \caption{The setup that was simulated to obtain ion and electron fluxes to the substrate.  
  \label{fig:numerical_setup}}
\end{figure}

The setup shown in fig.~\ref{fig:numerical_setup} was simulated to obtain ion and electron fluxes and spectra. 
This setup mimics setups used in the real experiments \cite{sheridan_1992,flanagan_2006,wang_2016}.
A 5~cm high and 10~cm in diameter cylindrical metallic chamber with grounded walls 
was filled by hydrogen with pressure from 0 to 10 Pa.
The bottom wall of the chamber was covered by a silicon dioxide substrate.
A 3~cm in diameter grounded grid was set in the center of the top wall.
The grid transparency was 75\%. 
An electrode placed in front of the grid emitted $2\cdot 10^{13}\ cm^{-2}s^{-1}$ electrons.
The potential applied to this electrode (-70~V) pushed these electrons inside the chamber.
The electrons ionized hydrogen and caused plasma formation.
The resolution of computational mesh was chosen to resolve the Debye length.

The simulation lasted until density and potential of the plasma in the center of the chamber 
and substrate potential reached a stationary value.
It took about several milliseconds of simulation internal time.
%The stationary potential of the dielectric substrate was about minus several Volts.
The dependencies of plasma density and fluxes to the substrate on hydrogen pressure 
are shown in fig.~\ref{fig:plasma_density}.
Stationary values are given.
Our model takes into account secondary electron emission (see description in sec.~\ref{sec:buildup}). 
The electron flux plotted in fig.~\ref{fig:plasma_density} includes only incoming electrons 
and does not include outcoming secondary electrons.
The last is the reason of the difference between stationary electron 
and ion fluxes in fig.~\ref{fig:plasma_density}.

\begin{figure}[ht]
  \includegraphics[width=\linewidth]{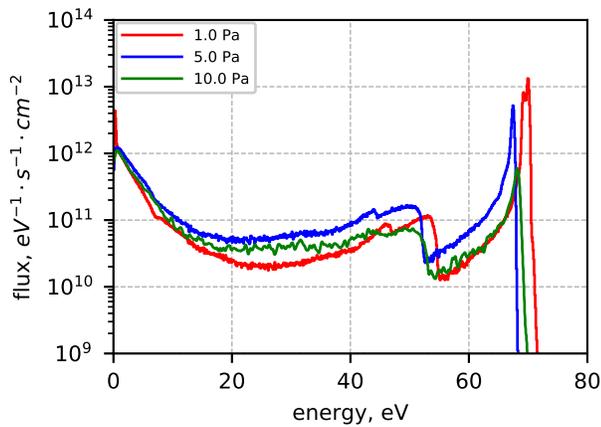}
  \caption{Spectra of electron flux to the substrate at different hydrogen pressures.  
  \label{fig:e_spectra}}
\end{figure}

Energy distribution functions of electron flux to the substrate at different pressure 
are demonstrated in figs.~\ref{fig:e_spectra}.
Every spectrum consists of two main components: 70~eV electron injected in the chamber 
and low-energy electrons of the formed plasma.
The peak of 70~eV electrons decreases with hydrogen pressure,
because more electrons interacts with the background gas molecules.
The shift of the right edge of the spectra with hydrogen pressure is caused 
by different stationary potential of the dielectric substrate at different gas pressure. 
It varies from -2.5~V to -0.5~V.
The average energy of electron flux reduces from 70~eV for 0~Pa to about 20~eV for 10~Pa of 
the background gas pressure.

\begin{figure}[ht]
  \includegraphics[width=\linewidth]{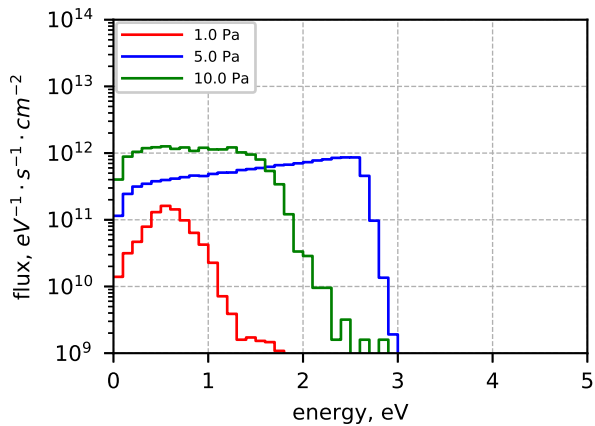}
  \caption{Spectra of ion flux to the substrate at different hydrogen pressures.  
  \label{fig:i_spectra}}
\end{figure}

Collisions between electrons and gas molecules produces $H_2^+$ ions,
that converts in $H_3^+$ ion in collisions with gas molecules within several microseconds.
Hence, $H_3^+$ is the main ion of the plasma.
Spectra of $H_3^+$ flux to the substrate at different gas pressures are given in fig.~\ref{fig:i_spectra}.
The energy of spectrum maximum equals to the plasma sheath voltage
as well as for electrons and varies with the background gas pressure.

\section{Simulation of particle charge buildup\label{sec:buildup}}
Ion and electron fluxes obtained at the first stage of the simulation were used to 
expose a dielectric particle resting on a dielectric substrate to calculate the force.
This simulation follows the idea of the simulation we performed earlier \cite{krainov_2020a} 
for an electron beam.
A brief description is given below.
A 100~nm dielectric particles was placed 1~nm above a dielectric substrate.
Two dielectric substrate of different thickness were used: thick (2000~nm) and thin (3~nm).
Ions and electrons bombarded the particle and the substrate.

The dielectric substrate rested on a grounded electrode,
that was the bottom boundary of the simulation domain.
The other boundaries were several micrometers away from the particle 
as the field of the charged particle is negligible at such a distance.
The top boundary was a biased electrode
Its potential was equal to the potential of the dielectric substrate from the first stage.
The side boundary was set to be a mirror.
It imitated the rest of the charged dielectric substrate.
A non-uniform rectangular grid was used in the simulation.
Its maximum resolution was equal to 1~nm to resolve the distance between the particle and the substrate.

The material of the dielectric we used is silicon dioxide ($SiO_2$).
We took into account dielectric permittivity ($\varepsilon = 3.9$),
electron induced secondary electron emission yield \cite{dunaevsky_2003},
the spectrum\cite{schreiber_2002} and angular distribution\cite{bundaleski_2015} of 
secondary electron emission.
The dielectric was assumed to be an ideal insulator.

\begin{figure*}[ht]
  \begin{minipage}[h][55mm][t]{0.49\linewidth}
  \center{\includegraphics[width=\linewidth]{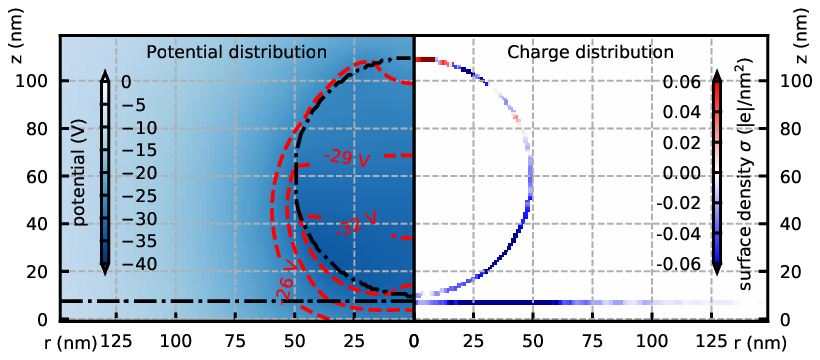}}
  \end{minipage}
  \hfill
  \begin{minipage}[h][55mm][t]{0.49\linewidth}
  \center{\includegraphics[width=\linewidth]{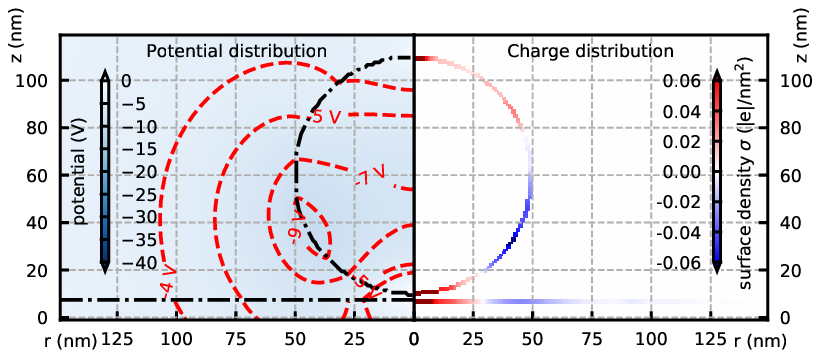}}
  \end{minipage}
  \caption{Distribution of accumulated charges over the particle and substrate surfaces for 
  two values of hydrogen pressure: 0~Pa (left) and 2.5~Pa (right). 
  \label{fig:dielectric_charge_density}}
\end{figure*}

\begin{figure}[ht]
  \includegraphics[width=\linewidth]{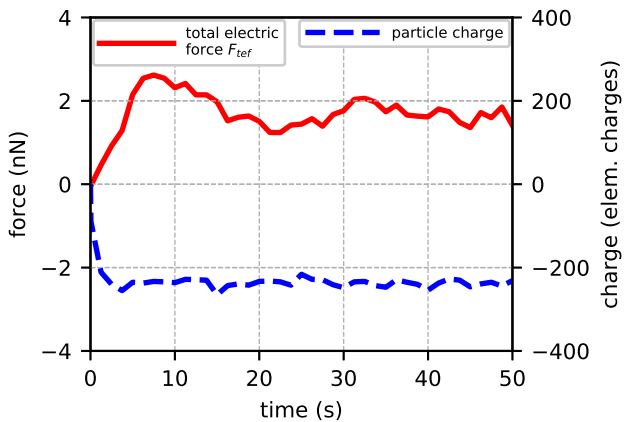}
  \caption{Dynamics of the particle charge and the electrostatic force acting on the particle 
  in case of 2.5~Pa of hydrogen pressure.\label{fig:force_dynamics}}
\end{figure}

The simulation lasted until the distribution of charges over the particle 
and the substrate surfaces reached a stationary state.
The distributions obtained in the simulations at two hydrogen pressures (0 and 2.5~Pa) 
are given in fig.~\ref{fig:dielectric_charge_density}.
The typical force dynamics obtained in case of simultaneous exposure to electron beam and plasma
is given in fig.~\ref{fig:force_dynamics}.

The left distribution in fig.~\ref{fig:dielectric_charge_density} 
corresponds to the exposure to an electron beam only.
The lower part of the particle accumulates secondary electrons 
emitted by nearby region of the substrate due to secondary electron emission.
This simulation was described previously in detail\cite{krainov_2020a}.
In case of plasma addition (the right distribution in fig.~\ref{fig:dielectric_charge_density}), 
the negatively charged lower part of the particle attracts slow ions 
(their energy is about several eV) pulls them in in the gap between the particle and the substrate.
This reason causes the formation of positively charged regions on the substrate and the particle 
in the place of contact.
In other words, the negatively charged particle acts as a focusing lens for ions.
In case of plasma addition, the upper part of the particle is charged positively too.
It directly experiences the bombardment of ions.
Besides, it experiences the bombardment of electron with energy larger 
than the first crossover energy for silicon dioxide.
Secondary electron emission yield is more than 1 for such electrons, 
i.e. in average such electrons leave a positive charge on a dielectric after collision.

Fig.~\ref{fig:force_dynamics} demonstrates the result of the simulation for 
2.5~Pa of hydrogen pressure.
The force shown includes both electrostatic components:
the repulsive interaction between charges accumulated on the substrate and the particle
and the mirror force.
The force was calculated as integral over particle of product of electric field and charge density. 

A charge distribution over surfaces obtained in a simulation is fluctuating.
Our 2-D PIC model can not calculate properly the force fluctuation caused by fluctuation of charge distribution
due to the axial symmetry of the model.
That is why a charge distribution was smoothed at the stage of postprocesing
to take into consideration the "macro" effects of electron and ion accumulation.
We used Savitzky-Golay filter to approximate the integral with variable upper limit 
over the particle and the substrate surface.
The order of the filter was equal to 3, because the function to be approximated has 2 extremum points.
The width of the filter was equal to one third of the corresponding coordinate 
(radius for the substrate and azimuthal angle for the particle) range.
The last thing to be noted is that fig.~\ref{fig:dielectric_charge_density} gives smoothed distributions.
Dynamics of the force and the charge at other pressures is the same as in case of 2.5~Pa hydrogen pressure.

\begin{figure}[ht]
  \includegraphics[width=\linewidth]{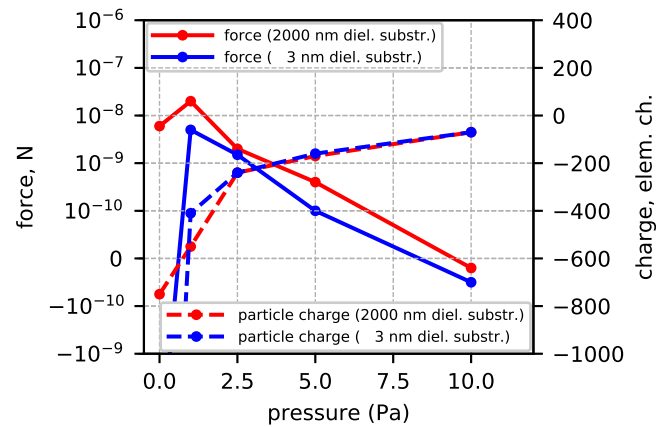}
  \caption{Dependency of the force acting on the particle and the particle charge obtained 
  in the simulations with two substrates of different thickness.
  \label{fig:force_vs_pressure}}
\end{figure}

Stationary values of the force acting on the dielectric particle and the particle charge 
for different hydrogen pressures are given in fig.~\ref{fig:force_vs_pressure}.
The most interesting feature of this plot is the transition from 0~Pa to 2.5~Pa,
i.e., plasma addition.
According to the conducted simulation, 
it is possible to take off a particle from a thick dielectric substrate with the help of 
a low-energy electron beam, but it is not possible in case of thin dielectric.
The plasma addition helps to cope with this problem.
This can be explained by two reasons.
Ion accumulation under the particle gives a significant contribution to the repulsion 
between the particle and the substrate.
Plasma addition results to lower total particle charge,
that decreases the mirror force.

\section{Discussion\label{sec:discussion}}
Table~\ref{tab:comparison} demonstrates a qualitative agreement between 
the results of our simulation and the published experiments.
To simulate a particle exposure to only plasma we exploited the same setup.
We registered ion and electron fluxes to the dielectric substrate in the region, 
that was not directly exposed to 70~eV electrons.

\begin{table*}[ht]
\caption{Qualitative comparison of the simulation results with the experimental results.
"+" sign means detected particle lofting, "-" sign means that particle lofting was not observed. 
The number in parentheses  is $l/\lambda$, where l is the height of the simulated setup (at the first stage),
$\lambda$ is the mean free path of an electron in hydrogen before ionization collision. 
This value reflects the portion of electrons absorbed by hydrogen and turned into plasma\label{tab:comparison}}
\begin{ruledtabular}
\begin{tabular}{*{9}{c}}
   & & electron beam & \multicolumn{4}{c}{electron beam + plasma} & \multicolumn{2}{c}{plasma (out of beam)}\\  
   & & 0 Pa (0) & 1 Pa (0.1) & 2.5 Pa (0.25) & 5 Pa (0.5) & 10 Pa (1.0) & 1 Pa & 10 Pa \\
  \hline
 \multirow{2}{*}{simulation} & thick (2000 nm) & + & + & + & + & - & - & - \\
  & thin (3 nm) & - & + & + & - & - & - & - \\
 \hline\hline
& & e beam & \multicolumn{4}{c}{e beam + plasma} & \multicolumn{2}{c}{plasma}\\  
 \hline
 \multirow{3}{*}{experiment} & Flanagan \cite{flanagan_2006} (thick diel.) & - & \multicolumn{4}{c}{+ (0.05)} & \multicolumn{2}{c}{-} \\
 & Sheridan \cite{sheridan_1992} (thick diel.) & - & \multicolumn{4}{c}{+ (0.1)} & \multicolumn{2}{c}{-} \\
 & Wang \cite{wang_2016} (partilces heap) & + & \multicolumn{4}{c}{+ (0.1)} & \multicolumn{2}{c}{-} \\
 
\end{tabular}
\end{ruledtabular}
\end{table*}

\section{Conclusion\label{sec:conclusion}}
The authors investigated a 100~nm silicon dioxide particle resting on a silicon dioxide substrate.
With the help of particle-in-cell numerical model, 
we exposed the particle to the flux of a low-energy electron beam, 
electron and ion fluxes of cold plasma and fluxes from the combination of these two sources.
We found that a plasma addition to an electron beam results to the ion accumulation 
between the substrate and the particle.
As a result, this plasma addition increases the force acting on the particle 
and make particle lofting possible from the substrate as thin as a naturally formed oxide layer.
The simulation results are in qualitative agreement with the published experiments.

\bibliography{particle-lofting-by-plasma-en.bib}

\end{document}